\documentclass{article}

\usepackage[utf8]{inputenc}
\usepackage[T1]{fontenc}
\usepackage{amsmath,amsfonts}
\usepackage{graphicx}
\usepackage[numbers]{natbib}   % numeric citations from \cite{}
\usepackage{hyperref}
\usepackage{url}
\usepackage{booktabs}
\usepackage{nicefrac}
\usepackage{microtype}
\usepackage[labelfont=bf]{caption}
\usepackage{xcolor}

\title{TerraTrace: Spatio-Temporal Vegetation Signatures for Land Use Analytics}

\author{
   \makebox[\textwidth][c]{Angela Busheska$^{a}$, Vikram Iyer$^{c}$, Bruno Silva$^{b}$, Peder Olsen$^{b}$,}\\
   \makebox[\textwidth][c]{Ranveer Chandra$^{b}$, Vaishnavi Ranganathan$^{b}$} \\
   \\
   $^{a}$Lafayette College, Easton, PA \\
   $^{b}$Microsoft Research, Redmond, USA\\
   $^{c}$University of Washington, Seattle, USA
}

\begin{document}

\maketitle

\begin{abstract}
Understanding land use over time is critical to tracking events related to climate change, like deforestation. However, satellite-based remote sensing tools which are used for monitoring struggle to differentiate vegetation types in farms and orchards from forests. We observe that metrics such as the Normalized Difference Vegetation Index (NDVI), based on plant photosynthesis, have unique temporal signatures that reflect agricultural practices and seasonal cycles. We analyze yearly NDVI changes on 20 farms for 10 unique crops. Initial results show that NDVI curves are coherent with agricultural practices, are unique to each crop, consistent globally, and can differentiate farms from forests. We develop a novel longitudinal NDVI dataset for the state of California from 2020-2023 with 500~m resolution and over 70 million points. We use this to develop the TerraTrace platform, an end-to-end analytic tool that classifies land use using NDVI signatures and allows users to query the system through an LLM chatbot and graphical interface.
%This work presents the idea to compute yearly spatio-temporal vegetation signatures from Spectral Indices measured through satellite imaging and to perform multi-year analysis based on them to estimate land-use. We further demonstrate how such cross-year signatures can be used to infer other information about a given geolocation, such as deforestation, agricultural crop growth, land degradation and wildfire risk. This workshop paper presents an end-to-end system which demonstrates the ability to leverage the Normalized Difference Vegetation Index(NDVI) signature curves to distinguish crops. Lastly, TerraTrace also shows the benefits of using a math-based Index computation with GPT's analytical skills to obtain land use inference, estimate risks and verify them with other modalities and data sources.
\end{abstract}
\section{Introduction}
\begin{figure}[t]
    %\hspace*{-1.0cm} 
    \centering
    \includegraphics[width=\linewidth]{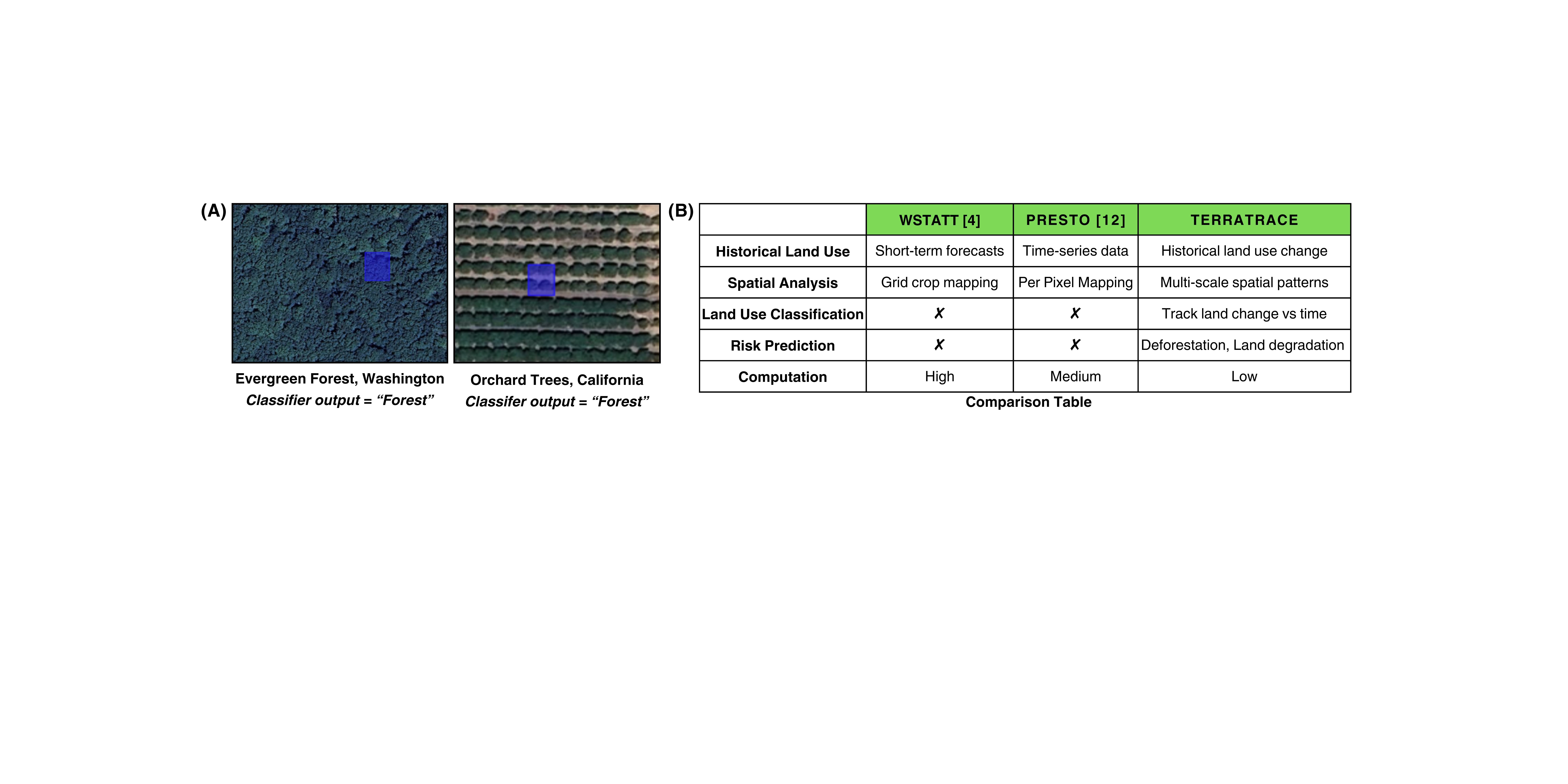}
    \vspace{-0.6cm} 
    \caption{\textbf{Land use classification challenges} (A) Current forest probability maps based on prior datasets~\cite{ALOS-paper, pittman2019global} cannot distinguish between farms and wild forests. (B) Unlike prior work \cite{presto} \cite{wstatt}, TerraTrace tracks historical land use with low computational overhead and variable scales.}
    \label{fig:Fig1}
    \vspace{-0.3cm} 
\end{figure}

% @Vaishnavi- we 
%As wildfires and natural disasters become more frequent due to climate change, maintaining an accurate history of land use over time is essential for food security and agricultural planning. Land use classification can help track changes in specific areas, providing critical data to understand prior land degradation and predict future risks. One such issue is deforestation. 
Every year humanity clears 10 million hectares of forests which releases over 5.6 billion tonnes of greenhouse gases annually ~\cite{UNEP2024}. This significant contribution to climate change has prompted passage of the European Union Deforestation Regulation (EUDR) ~\cite{EUDR}, which aims to ensure that products linked to deforestation are excluded from the EU market from December 2024. There is now a critical need for satellite and remote sensing technologies to help implement such regulations globally. However, these systems frequently misclassify vegetation types with similar visual characteristics. For instance, the Advanced Land Observation Satellite (ALOS) Forest Map is widely used to monitor forest growth and track deforestation \cite{ALOS}, but it falsely classifies orchards and pine plantations, as forests, as seen in Fig.\ref{fig:Fig1}(A). Similarly Fig~\ref{fig:Fig1}(B) shows that prior works do not provide data on historical land use changes at variable spatial scale which is needed to answer questions about deforestation. Additionally, they require substantial computing resources.~\cite{tseng2023lightweight, ravirathinam2024combining}.

%A comprehensive understanding of historical land use patterns is essential for accurate classification. To better understand land use across the years, a range of factors, including spatial considerations (e.g., topography), temporal variables (e.g., seasonal changes), agricultural practices (e.g., crop rotations and growth cycles), and risk assessments related to deforestation and land degradation need to be well understood.

To address these challenges we present TerraTrace: a temporal signature mapping system that combines Spectral Vegetation Indices, Satellite Imagery and Crop-Data Layer (CDL) \cite{montero,CropScape} to measure historical land use. The key insight here is that temporal variables and agricultural practices like crop rotations and growth cycles of plants are visible in satellite-based spectral index data. Specifically, we show that yearly patterns of Normalized Difference Vegetation Index (NDVI), based on plant photosynthesis, can distinguish wild forests from crops. We further find that these NDVI signatures are unique to different crops and follow consistent patterns throughout the world. Leveraging this we make the following contributions: \\
\textbf{1) Exploring NDVI Signatures.} We analyze yearly NDVI change on 20 farms with 10 unique crops. We verify the curves are consistent with agricultural practices, can distinguish between unique crops and forests. We also show they are consistent globally on coffee farms in Vietnam and Honduras. \\
\textbf{2) NDVI Dataset.} We develop a novel NDVI dataset for the state of California from 2020-2023 with 500~m resolution containing over 70 million points. \\
\textbf{3) TerraTrace Platform.} We develop the TerraTrace platform, an end-to-end analysis tool that classifies land use using mathematical analysis on NDVI signatures and allows users to query the system through an LLM chatbot interface and graphical interface.

%\textcolor{red}{TODO: Add point aout NDVI dataset - how many points for california, the fact that it doesn't exist but needs to be created}

%system with a GUI allowing users to input a set of coordinates  can take coordinates and provide detailed insights on land use.

%We develop and demonstrate the capabilities of TerraTrace for Agricultural land use in the state of California. Lastly, we demonstrate the possibility of using signature curves to identify the land use at a global scale. We display this through the example of coffee cultivation across three distinct regions, Guatemala, Vietnam and Honduras, all of which exhibit similar curves which correspond to coffee as shown in Fig.\ref{fig:Fig2}(C). We also discuss the possibility of scaling TerraTrace with multi-modal inputs and tools like SATClip \cite{klemmer2024satclip} to understand the geolocation-based features.
\begin{figure}[t]
    %\hspace*{-2.8cm} 
    \includegraphics[width=\linewidth]{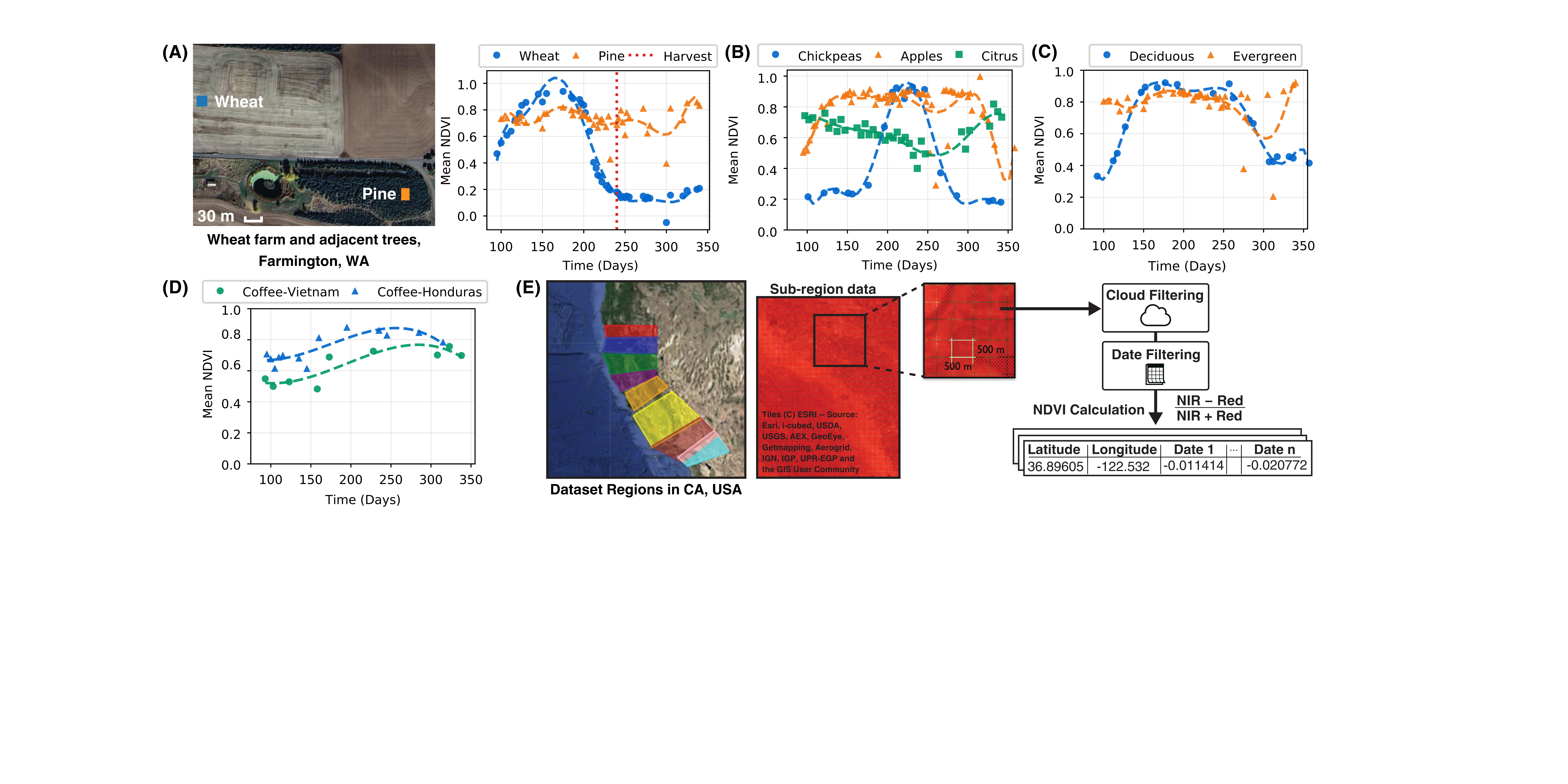}
    \vspace{-0.5cm}
    \caption{\textbf{NDVI Signature Curves.} (A) Comparison of  NDVI on a wheat farm versus adjacent pine trees in Farmington, WA from Apr-Dec 2020. (B) NDVI curves for a chickpea farm in Grand Rapids, ND, an apple orchard in Wenatchee, WA and a citrus farm in Fresno County, CA. (C) NDVI for deciduous forest in MO, USA versus evergreen forest in WA, USA. (D) NDVI data for coffee farms in Buon Ma Thuot, Vietnam and El Paraiso, Honduras. (E) NDVI dataset created for CA, USA.}
    \label{fig:Fig2}
    \vspace{-0.6cm}
\end{figure}

\section{NDVI Crop Signature Curves}
\label{signatures}
The NDVI metric is a widely utilized remote sensing vegetation indicator~\cite{USGS}. Molecules such as chlorophyll in live green plants have higher absorption for specific wavelengths of light (e.g. 400-500~nm, 600-700~nm) while longer near infrared (NIR) light not used in photosynthesis are reflected by the leaf cell structure to prevent overheating.%Live green plants appear darker in red versus NIR wavelengths~\cite{gates2012biophysical}. 
NDVI is calculated as a ratio of spectral reflectance measurements in the NIR and visible red bands:
\vspace{-0.10cm}
\begin{equation}\label{eq:1}
    \text{NDVI} = \frac{\text{NIR} - \text{Red}}{\text{NIR} + \text{Red}}
    \vspace{-0.25cm}
\end{equation}

Our key insight is that NDVI values vary uniquely over time based on growth cycles of specific vegetation and farming practices. To compute this unique ``signature curve,'' we first pre-process Sentinel-2 imagery from Google Earth Engine (GEE) and Microsoft Planetary Computers (MPC)~\cite{GEE, Sentinel}, removing those with a cloud mask ratio over 10\%. Next, we compute daily NDVI values at a 10-meter resolution for the target region and dates.  We shift the raw NDVI to be positive to exclude bare soil or clouds and normalize them to a 0-1 scale for consistent analysis across datasets.

We begin by analyzing data from April through December 2020. Fig~\ref{fig:Fig2}(A) shows an aerial image of a wheat farm in eastern Washington along with corresponding mean NDVI values over a year in a section of a wheat field and nearby stand of pine trees. The graph shows both the raw NDVI values as well as a degree eight polynomial fit line to illustrate the trend.

% What does this mean?
%The data undergoes non-linear least-squares minimization and curve-fitting procedures to refine these estimations. 

This wheat data shows a defined change in NDVI in key agricultural periods. After the spring wheat planting and germination in April, the NDVI values increase to a peak as the plants grow. Next, after day 200, we see a sharp drop for the pre-harvest dry down and harvest ending in late August. In contrast, the adjacent pine trees maintain a consistent high NDVI throughout the year. This data shows ability to distinguish crops from adjacent forested land.

Next we analyze 20 farms in the US and show examples in Fig\ref{fig:Fig2}(B). These include a chickpea farm in Grand Rapids, ND, USA an apple orchard in Wenatchee, WA, USA and a citrus farm in Fresno County, CA, USA. The data shows changes in NDVI that reflect agricultural practices. We also analyze forests as seen in Fig~\ref{fig:Fig2}(C). The decidious forest vegetation in Missouri, USA increases in spring and declines in fall unlike the consistent data from an evergreen forest in Washington, USA.

These curves also appear to be consistent globally, due to the similarity in growing cycles. Fig~\ref{fig:Fig2}(D) compares a coffee farm in Buon Ma Thuot, Vietnam and El Paraiso, Honduras. The data shows a similar trend in NDVI. We note these regions have less data available, and we adjust our curve estimate to use a degree three polynomial. Additionally there is a fixed offset which could be due to the difference in geography or specific crop variety.

\subsection{NDVI Dataset Preprocessing}
\label{dataset}
%VI Summary of changes: I tried to add a little more detail on how this works, some context for the scale of the dataset and reformatted it to remove the bullets.
To support applications from land use tracking to vegetation health we develop the first longitudinal NDVI dataset using Sentinel-2 data from 2020-2023. We select the state of California [423,970km$^2$] due to its agricultural diversity~\cite{USDA1}. To support small farms, we select a sub-kilometer grid scale of 500x500~m. This however results in almost 1.7 M locations, each of which requires multi-year time series data. To facilitate large datasets, we divide the state into eight regions as seen in Fig.~\ref{fig:Fig2}. 

We divide each region into a 0.25~km$^2$ grid and follow the procedure above to extract cloudless data. Next, we filter the date range to capture the full growing cycles of various crops and show vegetation dynamics. We then compute the NDVI with Eq.\ref{eq:1} using the Sentinal-2 B8 NIR band and B4 red bands. This 70 million point dataset provides a time series of vegetation health for each location and serves as the foundational input for generating signature curves for further analysis.

%Land use understanding
%spatial factors- topography, soil types
%temporal factors- seasonal changes, long term trends
%crop patterns- rotations, growth cycles
%risk assessmentdeforestation and land degradation

%Questions:
%1. What is the 

\begin{figure}[t]
    \centering
    \includegraphics[width=0.7\linewidth]{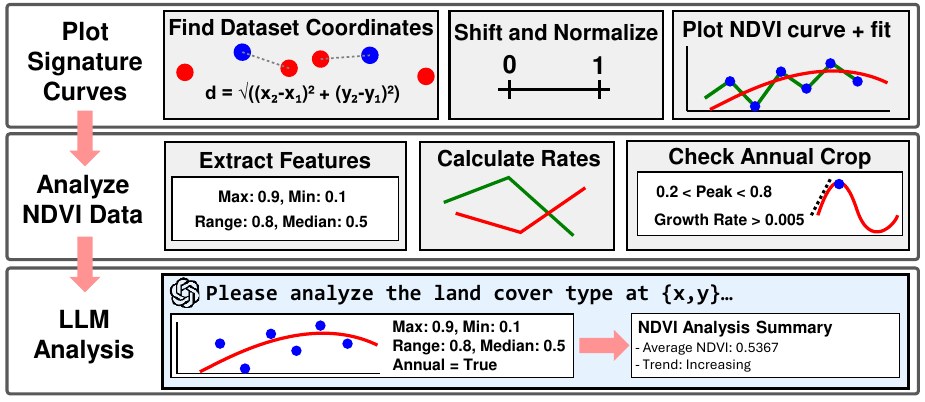}
    \vspace{-0.2cm}
    \caption{\textbf{TerraTrace System.} TerraTrace finds the NDVI coordinates from our dataset, extracts a set of metrics to analyze the data, and passes these to GPT-4 Turbo for additional analysis.}
    \label{fig:Fig3}
    \vspace{-0.5cm}
\end{figure}
\section{TerraTrace Platform}
\label{system}
We combine our insights about NDVI signatures from the dataset developed in Sec.~\ref{signatures} to develop TerraTrace, an end-to-end AI powered land use analysis platform. TerraTrace takes in a set of geographic coordinates that define the target region. From the dataset, we filter coordinates within this geo-polygon by coarse latitude and longitude ranges to identify the dataset region. Next we calculate the Euclidean distance between a target coordinate and points in our dataset. We extract the corresponding signature curves within the prescribed polygon by computing the mean NDVI value per time point across valid coordinates. We then interpolate the NDVI values and fit a 3rd order polynomial. TerraTrace presents users with a GUI that plots the region using the Leaflet interactive map library which creates an interface to adjust the polygon and plot the NDVI signature curve. 

TerraTrace also analyzes the data to extract land use insights. First, we check that the data is valid and the curve has $>$10 points for robust classification. We then extract features such as the annual minimum and maximum NDVI, range, and median. We determine the growth and decline rates by calculating the maximum point difference in NDVI values. We then use these metrics to check if the curve is an annual crop. This means that the NDVI increases above 0.2 indicating healthy vegetation growth, reaches a peak between 0.2 and 0.8, followed by a decline. We check the growth and decline rates are $>$0.005 to help filter out perennial species which only have small NDVI fluctuations across seasons.

TerraTrace complements these metrics with an LLM based analysis. We pass in the statistics such as max, min and average, an image of the NDVI curve, and a classification of whether the region contains vegatation using a JSON format. We determine vegetation presence with thresholds: <0.1 is non-vegetative, 0.1-0.2 as some vegetation, and 0.2 as healthy vegetation \cite{eos}. We pass in images converted to grayscale, resized, and encoded as a base64 string within the JSON. Next we construct prompts like the following to query the model. "The area of interest is defined by the $[{coordinates}]$. Please analyze the land cover type at this location." We use GPT-4 Turbo (version 2024-04-09) which translates the curves and data into a detailed analysis table, providing an additional validation. Further explanation of the algorithm is presented in Fig.~\ref{fig:Fig3}.

%We integrate additional data sources as well. We use the CDL to calculate a percentage of total crop pixels of specific types within the target region and 

%\textcolor{red}{TODO: move Historic Wild-Fire Data to the end}

%We pass in the statistics such as max, min and average, an image of the NDVI curve, and a classification of whether the region contains vegatation. We determine this using a simple threshold defining <0.1 as non-vegetative, 0-0.2 as some vegetation, and 0.2 as healthy vegetation. We pass in images converted to grayscale, resized, and encoded as a base64 string within the JSON. Next we construct prompts like the following to query the model. "The coordinates for the area of interest are {coordinates}. Please analyze the land cover type at this location."

%VI: I'm removing this because I don't know if it's clear that the compute required to serve an LLM is lower than this. Maybe it is but without hard numbers I think it may raise questions.
%After the curves are plotted, a comprehensive data analysis is conducted. We reduce the platform's dependence on extensive data and compute power by combining mathematical modeling with LLM analysis.

%To enhance the robustness of our analysis, we employ a Large Language Model GPT-4 Turbo. Utilizing up to 2000 tokens per query, the LLM translates the visual representation of the curves into a detailed analysis table, providing an additional layer of validation. Further explanation of the algorithm is represent in Fig.~\ref{Fig3}.

\section{Initial Demonstration and Results}
We perform an initial evaluation to show TerraTrace's potential. We input coordinates from a corn farm in Stockton, CA. Preliminary analysis using features extracted from the NDVI curve show the crop has expected growth patterns, notably an initial increase in vegetative indices during April and May, followed by a peak growing season spanning June, July, and August. The maximum value (a point in the growing season when the vegetation reaches its highest level) value of 0.97 and a median (measure of the overall vegetation health throughout the growing season) of 0.8722, which confirms the hypothesis of sustained healthy vegetation throughout the growing cycle. The loss rate of -0.7729 is consistent with the rapid dry-down phase in corn harvests. To validate our findings, we cross-referenced the results with Crop Data Layer information \cite{CropScape}, which confirmed that the coordinates indeed represent corn cultivation areas. We also load the MODIS dataset for the US, calculate distance between wildfires and the target location to determine the region's fire history. 

%Collectively, these metrics strongly suggest that the coordinates under investigation correspond to a specific agricultural land use type.

%Moved the EUDR use case discussion to the next section

\begin{figure}[t]
    %\hspace*{-2.5cm} 
    \includegraphics[width=\linewidth]{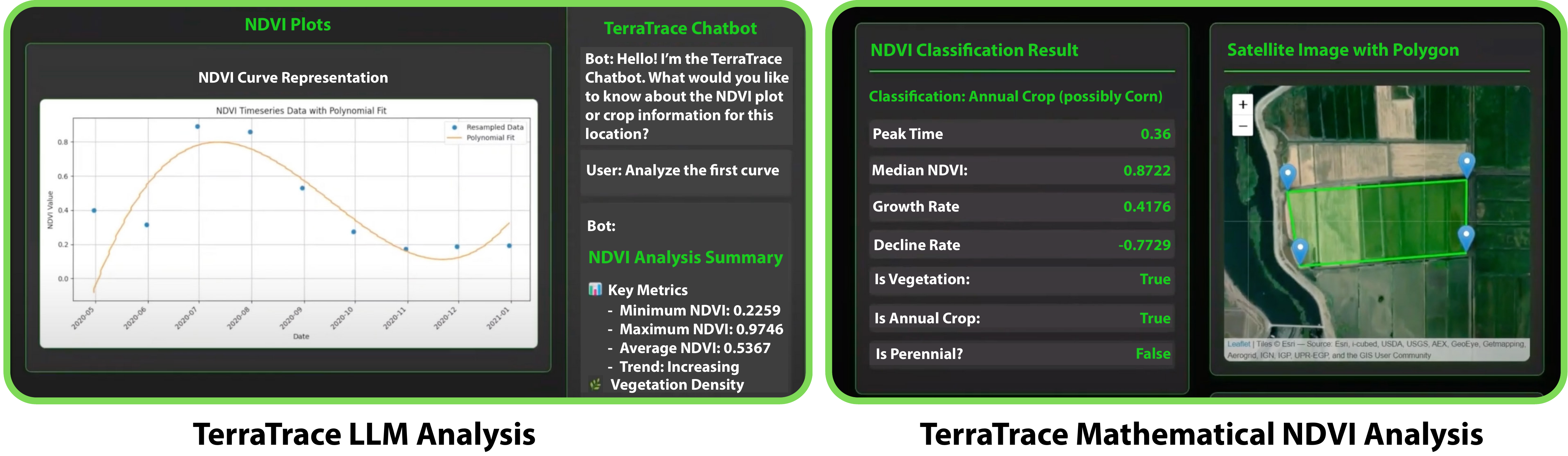}
    \vspace{-0.5cm}
    \caption{Screenshots of TerraTrace}
    \label{fig:Fig4}
    \vspace{-0.7cm}
\end{figure}
\section{Discussion and Future Work}

TerraTrace demonstrates that spectral signature curves can be used to understand land use across location and time. While we demonstrate this in a single state, the potential for it to scale globally can be seen in the example with coffee. In addition, while we only demonstrate the use of NDVI in this preliminary work, future versions of TerraTrace could include indices like Land Surface Temperature \cite{LST} and Irrigation Probability \cite{ML}. A full list of spectral indices can be found in \cite{montero}.  TerraTrace also faces limitations which need to be overcome to achieve accurate global land use estimation. The platform's geographic limitation to California necessitates dataset expansion. While aiming for global applicability, we recognize potential generalization challenges, suggesting the development of region-specific models \cite{klemmer2024satclip}. Spatial resolution of TerraTrace is also limited to 10m x 10m based on the Sentinel-2 capabilities. Temporal limitations due to infrequent satellite imagery could be mitigated through harmonized data collection every three days \cite{HLS}. Enhancements in spatial resolution, geographic coverage, and temporal frequency can improve TerraTrace's accuracy and global relevance in land use and climate change monitoring.

% Moved here from previous section
Incorporating these features could allow users such as EU regulators to track land use accurately over time, ensuring that deforestation-related decisions are based on data beyond paper certifications shared by suppliers. TerraTrace could enable small and large entities alike to automatically and objectively verify historical land use for compliance, thereby creating a fairer regulatory environment. In addition to regulatory compliance, we hope this approach can also be used for new applications; for example to aid farmers to project potential risks and towards climate change adaptation.
%On the other side, not all farmers have the time or resources to monitor whether their land has been affected by deforestation. TerraTrace helps them by providing accurate land use data \cite{IDH}.

%--------------------------------
% REFERENCES
%--------------------------------

\end{document}